\documentclass[pdflatex,sn-apa]{sn-jnl}

\usepackage{graphicx}%
\usepackage{multirow}%
\usepackage{amsmath,amssymb,amsfonts}%
\usepackage{amsthm}%
\usepackage{mathrsfs}%
\usepackage[title]{appendix}%
\usepackage{xcolor}%
\usepackage{textcomp}%
\usepackage{manyfoot}%
\usepackage{booktabs}%
\usepackage{array}%
\usepackage{placeins}%
\usepackage{csquotes}%
\usepackage{microtype}%
\setcitestyle{citesep={,}}
\patchcmd{\keywords}{\keywordname:}{\keywordname\enspace}{}{}

\AtBeginEnvironment{thebibliography}{\interlinepenalty=10000}

\begin{document}

\title[Early Epistemic Settlement]{Early Epistemic Settlement in AI-Assisted Writing}

\author{Han-yu Wang}
% Suppress the terminal period appended by sn-jnl after the email address.
\newcommand{\omitaffiliationperiod}[1]{}
\affil{The University of Hong Kong\par
\href{mailto:henry.why@connect.hku.hk}{\textcolor{black}{henry.why@connect.hku.hk}}\omitaffiliationperiod}

\abstract{In trying to complete a passage, an author can make connections among her materials that change what she can argue and the demands the argument must meet. A language model can supply a passage that does the work required at that point in her argument. Accepting it can end her own attempts before those connections have developed. I call this interruption early epistemic settlement. It can occur even when she fully understands the response and correctly judges it adequate for the passage's present role in the argument. The answer can satisfy the desire for resolution that kept her at work. Returning to her unfinished attempt would take more effort, and she may be unable to anticipate what she could achieve by continuing it. With repeated assistance, accepted answers shape what the writer asks next and which relations she goes on to develop. Useful answers can thus sustain inquiry while cutting short the work through which an author could form arguments that accommodate demands she has yet to recognize.}


\maketitle

\section{Introduction}

% P001
An author returns to her notes to qualify an explanatory claim. She needs to limit what she says to what her evidence supports. In trying to do so, she begins to work out how observations she had loosely grouped together belong in one explanation. Relating the observations may give her a way to explain them that she could not have formulated when she began. Accounts of writing as learning and thinking give such discoveries a central place \citep{Emig1977,Menary2007writing}. Now suppose a language model supplies a usable qualification while she is still trying to relate the observations. If she accepts it and moves on, what becomes of the work she had begun?

% P002
This interruption can occur even when the supplied passage is adequate and the author understands it fully. She can explain why the answer works and use it in the next part of her argument. Her own efforts might nevertheless have led her to form further relations among the observations, allowing her to organize the material differently and recognize further demands on the argument. She has yet to make these connections, so she may be unable to specify what continuing would add.

% P003
I use \emph{early epistemic settlement} for cases in which an adequate supplied answer interrupts the writer's constructive work. The answer settles the difficulty because it does what the passage needs to do at this point in the argument. Settlement is early in relation to the writer's unfinished attempt to organize her material. Accepting the answer lets her proceed without doing the further work through which new relations could have formed.

% P004
Writers select and evaluate material from model suggestions, sometimes changing their plans as they do so \citep{BhatEtAl2023}. In ``Reactive Writers'', engaging with suggestions displaced parts of the writers' own ideation. Participants described evaluation as less demanding than generating their own material, while often maintaining a sense of control over the result \citep{BhatEtAl2026}. The authors connect this pattern to satisficing, or accepting an idea once it seems good enough for the task. They show how suggested ideas continue to guide writers who actively revise the text.

% P005
In a visual ideation task, assistance from an AI image generator increased fixation on an initial example \citep{WadinambiarachchiEtAl2024}. \citet{QinEtAl2025} found that writers' ideas more closely resembled the model's suggestions when they used large language models from the beginning of ideation than when they first worked independently. In a study of strategic decision-making, assistance during problem formulation increased the number of alternatives while reducing strategic focus. The authors propose that early problem representations anchor later search \citep{WuKimLin2026}. Fixation can involve remaining with one idea when others are already available. Assistance can also interrupt an attempt through which another idea was becoming possible. The number of ideas in the resulting text does not show how the writer came to have them.

% P006
These findings raise a question for accounts of authorial control. \citet{Floridi2025} describes prompting and iterative refinement as a design practice called \emph{distant writing}. \citet{Rodrigues2026} locates authorial competence in the architectural, dialogical, evaluative, and integrative work of guiding a generative process. A writer who decides what her argument should establish is exercising this kind of control. We can nevertheless ask how she arrived at the aim she now pursues. Earlier responses may have helped determine that aim by resolving difficulties that were prompting her to reconsider what the argument should establish.

% P007
Several recent accounts address problem formation more directly. \citet{SabbahLi2026} examine how problems are discovered and redefined in human-AI collaboration. Their illustrative case treats acceptance of the first model-identified problem as satisficing and shows how further dialogue can revise the framing. \citet{CressKimmerle2026} argue that an apparently complete artifact can discourage inquiry before conceptual conflict emerges. In participatory problem structuring, \citet{BurgerWhite2026} argue that generated representations can stabilize meaning without the work through which participants would develop and contest it. These accounts make clear why supplying a representation of a problem may affect the formation of the problem itself.

% P007A
I argue that success in the immediate writing task can help bring this constructive work to an end. An adequate answer supplies the resolution the author sought. The desire that sustained her effort may be satisfied, while returning to her own attempt would take time and attention she no longer needs to spend to finish the passage. Such attempts can enable her to organize the argument in further ways and recognize demands absent from her initial task. She may leave the work before she can say what it would contribute. Repeated assistance carries the results of these settlements into later requests. I show how an inquiry can develop through the accepted answers while the relations that earlier attempts might have generated play no part in what the writer goes on to ask.

% P008
Section~2 explains how writers develop an argument by working out relations among their materials. Section~3 examines how an adequate supplied answer can bring this work to an end. Section~4 traces the relations that remain undeveloped in the case even when the author fully understands the supplied answer. Sections~5 and~6 examine what later inquiry builds on and the epistemic value of the work displaced by local success.

\section{How Writing Develops an Argument}

% P009
A writer has two intelligible passages in front of her, but each attempt to join them obscures the argument. In working on the transition, she realizes that they answer different questions. The problem is now how those questions belong in one argument. She had no set of candidate answers to this problem when she began trying to write the transition. She began with a difficulty in the text, and the effort to resolve it gave her a more definite problem to address. Formulating the question was itself an achievement of writing.

% P010
The idea that problems take shape during inquiry has a long history. \citet{Dewey1910,Dewey1938} follows reflective thought from a felt difficulty to the determination of a problem and describes inquiry as the transformation of an indeterminate situation. \citet{Schon1983} similarly treats problem setting as part of the practitioner's attempts to work out what sort of situation she faces. In writing research, \citet{FlowerHayes1980,FlowerHayes1981} examine the construction of rhetorical problems and the generation of goals during composition. Across these accounts, arriving at an answer depends partly on having learned what to ask.

% P012
Here the author is working out how two questions belong in one argument. Once she understands their relation, she can preserve it while trying different wording for the transition. Moving a sentence to another part of the argument may, conversely, let it answer a different objection or support a different inference even if its wording remains the same. Changing the words and changing what the claims do are distinguishable parts of the writer's work. As she moves between notes and draft, she is working out how the claims bear on one another as well as how to express them.

% P013
Consider a distinction the writer already understands in another context. In trying to place an objection, she discovers how to use the distinction in her draft. She has gained a use for something she knew. Drafting two claims may lead her to notice an assumption they share, and so to grasp a relation that understanding either claim on its own had not given her. In each case, the connection she has just made gives her a further way to work with familiar material.

% P017
Even after grasping such a relation, the writer may struggle to express it in a satisfactory transition. She can draw on it tacitly as she drafts and when she returns to the text. Help in finding the words can then allow her to express something she already understands. In the transition example above, however, the relation itself was still being worked out. Trying to write was changing how the author understood the claims to bear on one another.

% P011
A connection the writer has just grasped may have no obvious place in a passage that already serves its rhetorical purpose. To express it, she may have to change the order of claims or give a sentence a different job. \citet{GalbraithBaaijen2018} distinguish between the generation of content from implicit understanding and the organization of material to meet rhetorical goals. A passage that meets its rhetorical goals can thus leave the writer with something she has yet to articulate. Making room for this content can change the larger argument, since what she manages to express in this passage becomes material for the next.

% P014
% P015
Suppose the author is discussing a report of an observed regularity. Her draft attributes the regularity to a proposed mechanism, although the evidence establishes only the regularity. The next inference needs the observational result. If she qualifies the causal claim, she can keep that result as a warranted premise and proceed with the inference. I call the passage adequate if it meets these evidential and argumentative requirements. The qualification does the work needed at this point, even though further work on the material could lead the author to reconsider what the larger argument should establish.

% P016
To write the qualification herself, the author might try to say what the mechanism would explain if it were established. This takes her back to her notes, where she has grouped this report with others without yet working out how one explanation could account for their observations. What exactly is the mechanism supposed to explain about them? A distinction that seemed incidental when she made her notes may now prevent her from saying what she wants to say. As she tries to deal with it, she may find a different way to relate the observations, and she may come to require something of an explanation that she had not thought to ask when she began.

% P018
An author may become able to formulate a further explanatory proposal as she works out relations among her materials. At the outset, she may have had no such proposal to evaluate. \citet{KalisEtAl2013} draw a related distinction between generating possible actions and evaluating them, allowing generation to involve implicit goals and environmental prompts. The problem the author is trying to solve may change in the process. In an ill-structured problem, information from outside the initial problem space can help determine what the problem becomes \citep{Simon1973}. \citet{Boden2004} distinguishes exploration within a conceptual space from changes that make new kinds of ideas possible. In writing, the relations an author forms can enable her to construct an argument she could not have constructed with her earlier understanding. Developing it may then reveal requirements she had not previously recognized.

% P019
A source is organized around its author's questions, and a reader looking for something specific may encounter claims and distinctions that have no immediate place in her draft. Working out their bearing may force her to separate matters she had treated together, or to ask how apparently separate matters depend on one another. She can come away with a reason to revise the question that led her to the source. Rereading can have this effect too. Once the argument has changed, she approaches a familiar passage with difficulties that did not exist at the first reading, and may find a use for material that previously seemed irrelevant.

% P020
What the author can construct and what she asks of an argument need not change together. A new evidential requirement may force her to abandon an argument without giving her a replacement. Conversely, a connection between familiar claims may show her how to meet a requirement that has stayed the same. She may also come to see that one of her demands was mistaken and should be withdrawn. A passage can meet her current requirements while such work remains unfinished. Continuing may give her another way to resolve the difficulty, or lead her to revise what she takes a resolution to require.

\section{Early Epistemic Settlement}

% P021
Until she can make a passage work, the writer has a practical reason to keep thinking. Her attempts to connect its claims can take her into notes and sources whose relevance she is still trying to establish. An adequate response can give her a connection she understands and can use without resolving how that earlier material bears on her argument. If she adopts it, she can concentrate on fitting the answer into the draft and using the connection it supplies, leaving unfinished the attempt that was drawing her back to her notes.

% P022
Consider again the objection the writer could not place. Reading the model's suggestion, she sees which claim the objection challenges and why the reply belongs at that point in the draft. She can fully understand this connection and put it to use. Yet in trying to place the objection herself, she had been working through other claims whose bearing on it was unclear. As she worked out their relations, she might have found a different role for some of those claims. The suggestion resolves where to put the objection before she has done that work. Putting the objection in a suitable place can therefore end an attempt that might have led her to organize the argument differently.

% P024
Suppose the author remembers the passage in her notes that would not fit, or the connection she was trying to articulate. She knows where she could resume. What does that tell her about the value of doing so? If she had already worked out a further claim, she could ask whether it deserved a place in the argument. At the point considered here, working on the difficult material is still part of arriving at something she could assess in that way. Her memory gives her somewhere to begin without supplying the result. She can know where she left off and still lack a definite contribution to weigh against the effort of continuing.

% P025
% P026
Why would she stop while this work remains unfinished? The writer was looking for a resolution, and the model has given her one. \citet{KruglanskiWebster1996} describe a desire for definite knowledge that can motivate search before a judgment is reached and favor preserving it afterward. In this case, the same desire that kept the author at work can be satisfied by the answer. In reasoning experiments, lower feelings of rightness accompanied longer reconsideration and more frequent changes of answer \citep{ThompsonEtAl2011}. \citet{Nguyen2021} examines how felt clarity can serve as a signal to stop inquiry. Our writer can check the passage and find that it really does what she needs. This may reinforce her sense of resolution, even though she has left unresolved the relations she was trying to work out in her notes.

% P026A
\citet{CressKimmerle2026} explain how an apparently complete artifact can make discrepancies with the user's understanding less salient and so discourage engagement. Our author may understand the response well enough to find no such discrepancy. Rereading it or checking its claims against the available evidence can confirm that it does what the passage requires, while the relation of the material in her notes to the argument remains unresolved. The model can give her a usable passage without resolving the difficulties she encountered in her notes. Working through those difficulties could still give her something further to say, but she can finish checking the answer without examining them.

% P027
% P028
Before receiving the answer, the author was spending time on the material in order to finish the passage. She can now finish by accepting the response. Returning to the notes would take more time and attention, and she has no definite contribution yet to assess. \citet{LiederGriffiths2020} consider cognitive strategies in relation to their returns and computational costs. From the standpoint of completing this passage, accepting the answer can make good sense. She gets the result she needs and saves the effort of continuing an attempt whose outcome is uncertain. Stopping can be attractive on these grounds without requiring her to judge that the relations she has yet to form would be unimportant.

% P031
Accepting a result once it is good enough is familiar satisficing \citep{Simon1956}. To produce an adequate passage herself, the writer may have to form a relation between two observations that lets her meet an unchanged requirement. Work on other material may instead lead her to recognize that her initial requirements ask too little of the argument. A supplied passage can meet her current requirements before either development occurs. She can therefore stop because the passage is good enough, without having done the work through which either change might have arisen.

% P032
% P029
A useful source can leave the writer with just this sort of work to do. Its argument was developed around its author's questions, independently of the draft in front of her. To see how it bears on her draft, she may need to work through claims she had not intended to use. One of them may give her a reason to change what she was trying to say. A template likewise leaves her to decide how her particular material fits the proposed arrangement. In these situations, finding something useful begins work that is still needed to produce an adequate passage. Further search has a cost, and its expected value helps explain whether she continues looking for material \citep{PirolliCard1999}. A model can instead supply the connection that the draft needs in response to the request she is already able to make. This fit matters because it can spare her the encounters and attempts to connect material through which she might have changed the request itself.

% P032A
The request and the material supplied from the draft give the model a context in which to generate a response. Training with demonstrations and human feedback can improve its ability to follow the instructions it receives \citep{OuyangEtAl2022}. Consider what such a response does in the example from Section~2. It can remove the unsupported attribution of a cause and retain the observation as a premise for the next inference. The writer now has a way to use her evidence at this point in the argument. The passage works because it drops the explanatory claim that was taking her back to the reports. To supply this revision, the model need not resolve how those reports belong in an explanation.

% P033
A teacher who reads the draft could supply the same qualification. With a language model, the writer can obtain such help throughout composition, submitting the revised draft when the next difficulty arises. The draft now includes claims or qualifications accepted from earlier responses. When the model receives the revised draft, these claims and qualifications help determine what the next answer needs to do. Material that the writer was struggling to relate may have no place in this request, although working through it could have changed what she asked. She can obtain an answer that fits without returning to that material. The writer can keep resolving difficulties and advancing the argument along this route, without completing the earlier attempts that might have taken it elsewhere.

% P061
At the next step, the writer works with a claim she has obtained from the model. She may have to revise other claims to accommodate it or work out what follows from it. Cognitive offloading reduces a task's demands through external action \citep{RiskoGilbert2016}, and accounts of AI assistance examine how the user's activity changes. \citet{RiveraNovoaDuarteArias2026} distinguish assistance that complements the learner's activity from assistance that substitutes for it. The distinction between dependent and autonomous offloading concerns whether the user retains cognitive agency and uses assistance to support further thought \citep{ZhuEtAl2026}. The writer can exercise such agency in developing the supplied claim. But which of her earlier difficulties does this further work take up?

% P035
Research on \emph{integrative leaps} examines the invention involved in fitting a machine suggestion into a story \citep{SinghEtAl2023}. The writer creates connections that make the supplied material work in the story. To understand what this tells us about early settlement, we need to ask what connections the writer is making. Her earlier effort was directed by difficulties in material she already had. The suggestion gives her something else to develop, and making it useful can lead to connections that leave those difficulties untouched. We therefore need to trace the writing to see whether she is taking up that earlier attempt. The case in the next section follows such an attempt into the relations it might generate.

\section{What Local Success Can Leave Undeveloped}

% P037
The author in Section~2 has written, ``The proposed mechanism explains the recurrence.'' To qualify this sentence while retaining its explanatory aim, she tries to say what the mechanism would explain if it were confirmed. She consults the notes headed ``persistence through change,'' in which she brought together reports that seemed to describe one phenomenon. The heading now gives her trouble. What persists, and through what kind of change? Until she can answer that question, she cannot say what her proposed explanation is meant to cover. Revising the claim has taken her into the unfinished task of relating these reports.

% P040
While she is still working on the reports, she asks the model for a revision. It replies, ``The outcome recurred under the conditions reported. These observations do not establish the proposed mechanism.'' This agrees with the evidence in the report. The author understands that she can use the observation in the next inference even though its cause remains uncertain, and she accepts the revision. The model has removed the explanatory claim that was taking her back to the notes. She proceeds from the observation of recurrence, with the reports still grouped under their old heading and their relation unresolved.

% P038
Consider how her own attempt might continue. Some reports describe the same outcome recurring after a change in procedure. Another describes a process continuing after a change in procedure while its outcome changes. The heading ``persistence through change'' treats both as reports of one phenomenon. She tries, ``The mechanism would explain the recurrence described in these reports.'' This leaves out the report with the changed result. Substituting continuity for recurrence creates another problem. A similar outcome could be produced by a different process, so evidence of recurrence alone does not establish that the process has continued. Neither formulation says what she took the reports to share when she grouped them. If she is to keep them together, she must work out how continuity of a process might be related to recurrence of an outcome.

% P039
A continuing process may change how it operates. Perhaps the outcome recurs after a change in procedure because the process adjusts to that change. The writer now has a possible connection between continuity and recurrence to develop. To explain recurrence in these terms, she would need evidence of what changes within the process and how those changes preserve the outcome. She can bring the report with the changed result into the same inquiry, since it may show circumstances in which the adjustment fails to preserve the outcome. Details she did not need to check the model's qualification could now matter to an explanation of stable and changed results together. Following the proposal would take her back to the reports with a further purpose, to investigate how the process responds to different procedures.

% P041
The writer began by trying to qualify a claim about recurrence. She can now pursue an explanation of recurrence and variation in terms of how a process adjusts. Her heading had not supplied this relation between adjustment and stability, and she had no such proposal to evaluate when she began. Her attempts have given her an explanatory proposal she can now evaluate. Developing it would require her to account for stable and changed outcomes together. She had understood the reports before she could formulate this demand, which emerged as she worked out how they might belong in one explanation. The work has given her something further to propose and, with it, a reason to ask more of an explanation than her initial task required.

% P042
The difference begins with what the qualification has to preserve. In her own attempt, the author keeps the explanatory aim and has to work out how the reports belong together. The model's revision withdraws that aim from the passage. It is adequate because the next inference needs only the observation, for which there is already evidence. The writer can therefore go on with the draft before her attempt to relate the reports has yielded an explanation.

% P044A
We can discuss what this explanation might contribute because we have followed the attempt beyond the point at which the author stopped. She had yet to form the connection between adjustment and stability, so she could not assess it when she accepted the qualification. She may remember the troublesome grouping and know where she could resume. What she remembers has yet to give her the proposal we can now discuss. At the stopping point, she can therefore identify unfinished work without being able to identify this possible result of continuing it.

% P045
The qualification could also enter the draft without the author understanding why it works. She can retain the observation as a premise and carry over the qualification of the cause along with it. Her text would rely on a distinction that she has yet to grasp. Accounts of extended and distributed cognition examine how external representations participate in cognitive activity \citep{ClarkChalmers1998,Hutchins1995,Menary2007}. \citet{Heersmink2015} distinguishes using an external resource fluently from understanding the information it carries.

% P046
Changing the qualification to ``It remains unclear whether the outcome recurred'' would put the observation itself in doubt. This revision withholds a premise the report supports and the next inference needs. An author who understands why the original qualification worked can reject this change and retain the observation while leaving its cause unsettled. An author who has retained only the wording may accept the revision as another cautious formulation and remove the premise she needs without noticing the shift from explanation to observation.

% P047
Rejecting this revision need not involve giving a general explanation of the distinction. The writer may see that the new wording wrongly puts recurrence in doubt and preserve the observed result, even though she struggles to say why the cause is a separate matter. Her response would show a grasp of the relation that is still tacit. Repeating the model's original sentence, by itself, would not show this.

% P047A
A writer who understands the distinction can use it to connect a new report to her argument. The findings may strengthen the observational claim, or change what she needs to establish about the proposed cause. Research on self-explanation examines this relation between understanding and later use. \citet{ChiEtAl1989} found that students who related solution steps to principles in their explanations developed knowledge they could use when solving later problems, with less reliance on the examples. If the writer has only retained the wording of the qualification, she has yet to grasp the distinction and may miss how the report changes the support for an observation or the requirements for explaining it.

% P047B
Now suppose the model supplies the passage connecting this new report to the argument. The distinction between an observation and its proposed explanation can again do its job in the text even if the author still does not understand it. A difficulty that might have led her to examine it has been resolved for her. Here the earlier interruption is compounded. The first answer interrupted work that might have led her to connect adjustment with stability. Failing to understand the answer leaves her without a further relation, the distinction that could help her see how later reports bear on the argument. The next supplied connection can let her proceed without acquiring this resource for further thought.

% P049
Return now to the author who understands the qualification and rejects the mistaken revision. She can assess the evidence and use the distinction between the observation and its cause in further reasoning. \citet{MesseriCrockett2024} examine illusions about both the depth of understanding and the breadth of exploration. Early settlement need not involve either misjudgment or the loss of cognitive skills that concerns \citet{Heersmink2024}. The author's unfinished attempt was to say how reports of recurrence and continuity belonged together. Understanding the supplied qualification does not give her the relation between adjustment and stability that this work might have produced. She can use the answer competently while leaving unformed a further argument that her own attempts might have made possible.

\section{How Supplied Answers Shape Later Writing}

% P050
The writer has kept the observation of recurrence as a premise. She now returns to the notes to find other procedures under which the outcome recurred. Before she has worked these reports into the next passage, she sends the model her notes and the revised draft, asking it to identify the relevant procedures. The answer brings together the observations supporting recurrence under each one. It leaves out the report of a continuing process with a changed result, since the report contains no instance of the outcome she has asked about. She understands why it is absent and accepts the broader recurrence claim, now supported by observations she had not used in the first passage.

% P051
The changed result had mattered precisely because it combined a continuing process with a different outcome. Under her earlier heading, she had to say what it shared with reports in which the outcome recurred. The second request imposes no such demand. Once she is looking for further instances of recurrence, she can put that report aside without resolving its place in the grouping. She has therefore returned to the notes for a different purpose. Her earlier attempt to relate the reports can remain unfinished even as she draws more of them into the argument.

% P055A
If she then tried to explain this broader finding, an explanation of the recurrence originally reported might leave the additional instances unexplained. Asking it to cover these additional instances would enlarge the explanatory task, without yet asking how a continuing process can produce both stable and changed outcomes. Neither new content nor a demand that an explanation cover more would, by itself, show that she had resumed the earlier attempt.

% P055B
The proposal about adjustment would change the question. She could then ask how a process preserves its results under some changes in procedure and fails to do so under others. The author had no such proposal when she accepted the first answer. In trying to articulate her grouping, she might have formed it. Instead, she began with the recurrence claim she had accepted and asked which other reports could support it. The second request makes sense in light of what the first answer has settled. Its answer can be adequate for the request while the work that might have changed the request remains undone.

% P055C
\citet{BurgerWhite2026} describe how generated representations enter subsequent deliberation and are reinforced by their use. In participatory problem structuring, a representation can come to function as settled without the work needed to establish its legitimacy. Our writer can explain what the passages establish and how they support her next inference. In describing what she still needs to do, she can now begin with the accepted recurrence claim. She can identify the inferences she wants to develop from it and work through them. This description need no longer include her earlier difficulty in grouping the reports. With each further use of the accepted claim, she can leave the attempt further behind without recognizing that it might have changed which observations belonged in an explanation.

% P056
The participants in ``Reactive Writers'' often kept the framing of suggested ideas as they elaborated and reworded them \citep{BhatEtAl2026}. To study early settlement in this activity, a useful starting point would be the particular difficulty a writer was trying to resolve when she accepted a suggestion. Research on fixation and the timing of assistance gives us further reasons to attend to this moment in the development of an idea \citep{WadinambiarachchiEtAl2024,QinEtAl2025}. Successive drafts and model exchanges could show when the answer arrived and how the writer used it in later requests. Source-use records and the writer's accounts during composition could help reconstruct the attempt already under way. Process-tracing studies of AI-assisted second-language writing document substantial variation in how learners evaluate, revise, and incorporate suggestions \citep{AlghamdiAlghizzi2026}. Following a difficulty across these records would let researchers investigate whether the attempt continued after acceptance, and whether its developing relations contributed to what the author went on to write.

\section{The Epistemic Value of Writing}

% P058
An explanation of recurrence across procedures could leave the report with the changed result untouched. That report does not bear on whether the same outcome occurred under the procedures already listed. It becomes relevant once the writer asks how adjustments within a process sustain an outcome. A case where the result changes may then show the limits of those adjustments. The relation between adjustment and stability thus gives the excluded report a role in explaining the observations. What had fallen outside the recurrence claim can help determine the form of an explanation that covers recurrence and variation together.

% P059A
Suppose the writer develops the idea that adjustment preserves the outcome. Explaining stable results under each procedure separately would still leave her to say how the changed result fits the proposed relation between adjustment and stability. To account for both, she would have to show how adjustments within the process preserve the outcome under some changes in procedure and why they fail to preserve it under others. The changed result now bears on how recurrence itself can be explained. Having formed this relation, she now needs her explanation to meet this further demand. In meeting it, she could connect observations she had not previously brought into a common explanation, making an intellectual contribution through the effort to say how the reports belong together.

% P058A
Trying to relate the reports could also lead the author to revise the demand for a common explanation. The reports may turn out to concern different processes, so that bringing them under one explanation would misrepresent the observations. The investigation could then leave recurrence as the subject of the argument while giving her grounds for separating the reports. She would have learned why the broader grouping fails, and readers could assess her reasons for excluding observations that initially seemed relevant. In this case, the exploration would change which demands deserve to govern the argument. Better grounds for its scope would be an achievement of working through the material, even though the demand that initially prompted that work was withdrawn.

% P059
Developing the relation between adjustment and stability could give other researchers something to work with. They could investigate which changes in procedure the process accommodates while preserving the outcome, and where a different result should be expected. The writer would open up these questions through her effort to connect reports she already understood individually. This prospect gives that effort a value beyond her own learning. In an educational task, the understanding to be acquired may already be specified. Accounts of generative AI in education examine how obtaining a product can come apart from acquiring the epistemic abilities that the task is intended to cultivate \citep{Cassinadri2024,AylsworthCastro2024}. Acquiring those abilities is an achievement even when the argument is known. In the research case, forming the connection could also change what is available for others to explain and investigate. The work that remains unfinished may therefore matter to the inquiry's contribution as well as to its author's understanding.

% P065
% P030
Before the qualification arrived, the writer's effort to complete the passage was also an attempt to work out how the reports belonged together. With the supplied qualification in place, she can finish without pursuing their relation. Returning to the notes could still help her form an explanation, or discover why the reports should be treated separately. She can have this purpose in mind before she knows which relation the work might yield. She can now recognize that this exploration is worth pursuing for what it might contribute to the inquiry, even though completing the passage no longer requires it.

\section{Conclusion}

% P068
I have argued that accepting an adequate model answer can end work through which a writer might have developed a further argument. This interruption can occur even when she fully understands the answer and correctly judges that it does the work required in the passage. The answer can satisfy the desire for resolution that kept her thinking. Returning to her own attempt would take more time and effort, and she may be unable to anticipate what she could achieve by continuing it. What remains unfinished could have enabled her to form connections that change both the argument she can construct and the demands it should meet.

% P070
When the writer turns to the model again, the answers she has already accepted help shape her next steps. The revised draft supplies the context for the next request, and the next answer can develop claims already accepted in it. In our case, the writer moves from the finding in one report to a claim of recurrence under several procedures. An explanation of this broader finding would have to account for recurrence under the additional procedures. She has nevertheless left unfinished the attempt through which she might have connected adjustment with stability and explained recurrence and variation together. Developing that relation could have changed what she asked the explanation to cover. A succession of useful answers can therefore advance the inquiry without bringing such a contribution into view. Early epistemic settlement can affect the arguments an author becomes able to construct, including arguments that could meet demands her present questions have yet to reveal.

\bibliography{ref}

\end{document}